\documentclass[final]{IEEEtran}

\usepackage{balance}
\usepackage[latin1]{inputenc}
\usepackage{graphicx}
\usepackage{amssymb,amsmath,amsfonts,amsthm,mathtools}
\usepackage{algorithm,algorithmic}
\usepackage{cellspace}
\usepackage{cite}
\usepackage{float}
\usepackage{graphics}
\usepackage{makecell}
\usepackage{subfigure}
\usepackage{xspace}
\usepackage[usenames,dvipsnames]{color}
\usepackage{epsfig}
\usepackage{verbatim}
\usepackage{url,changebar,bm,xspace,dsfont}
\usepackage[T1]{fontenc}
\usepackage{setspace} 
\usepackage{soul} 
\usepackage{tabularx}
\usepackage{textcomp}

\def\BibTeX{{\rm B\kern-.05em{\sc i\kern-.025em b}\kern-.08em
    T\kern-.1667em\lower.7ex\hbox{E}\kern-.125emX}}

\def\SR{\mathop{\{S\hspace{-0.09cm} \rightarrow \hspace{-0.09cm} R\}}} 
\def\RD{\mathop{\{R\hspace{-0.09cm} \rightarrow \hspace{-0.09cm} D\}}}

\def\SD{\mathop{\{S\hspace{-0.09cm} \rightarrow \hspace{-0.09cm} D\}}}

\newcommand{\ie}{\textit{i}.\textit{e}., }

\floatstyle{ruled}
\newfloat{model}{H}{mod}
\floatname{model}{\footnotesize Model}
\newfloat{notatio}{H}{not}
\floatname{notatio}{\footnotesize Notation}

\newtheorem{remark}{Remark}

\newcommand{\NP}[1]{\textcolor{OliveGreen}{{#1}}}

\newenvironment{list4}{
  \begin{list}{$\bullet$}{
      \setlength{\itemsep}{0.05cm}
      \setlength{\labelsep}{0.2cm}
      \setlength{\labelwidth}{0.3cm}
      \setlength{\parsep}{0in}
      \setlength{\parskip}{0in}
      \setlength{\topsep}{0in}
      \setlength{\partopsep}{0in}
      \setlength{\leftmargin}{0.17in}}}
      {\end{list}}

\begin{document}

\title{On the Interplay Between Deadline-Constrained Traffic and the Number of Allowed Retransmissions in Random Access Networks}



\author{{Nikolaos Nomikos}, \IEEEmembership{Senior Member, IEEE}, {Themistoklis~Charalambous}, \IEEEmembership{Senior Member, IEEE}, \\{Yvonne-Anne Pignolet}, {Nikolaos Pappas}, \IEEEmembership{Senior Member, IEEE}
\thanks{
Preliminary results of this work appeared in \cite{Nomikos:2018SPAWC}. In this paper, we extend \cite{Nomikos:2018SPAWC} to include the case {of multi-packet reception}.

N. Nomikos is with the {Department of Port Management and Shipping, National and Kapodistrian University of Athens, 34400 Euboea, Greece (e-mail: {nomikosn@pms.uoa.gr}).}

T. Charalambous is {with the Department of Electrical and Computer Engineering, School of Engineering, University of Cyprus, Nicosia, Cyprus (e-mail: charalambous.themistoklis@ucy.ac.cy). He is also} with the {Department of Electrical Engineering and Automation, School of Electrical Engineering, Aalto University, Espoo, Finland (e-mail: {themistoklis.charalambous@aalto.fi}).}

{Y.-A. Pignolet is with Dfinity Foundation, Zurich, Switzerland (e-mail: {yvonneanne@pignolet.ch}).}

N. Pappas is with the {Department of Computer and Information Science, Link\"{o}ping University, Link\"{o}ping, Sweden (e-mail: {nikolaos.pappas@liu.se}).}
}
}

\maketitle

%
%
\begin{abstract}
In this paper, a network comprising wireless devices equipped with buffers transmitting deadline-constrained data packets over a slotted-ALOHA random-access channel is studied. Although communication {protocols facilitating} retransmissions increase reliability, packet transmission from the queue experiences delays and thus, packets {with time constraints} might be dropped before being successfully transmitted{, while at the same time causing the queue size of the buffer to increase}. {Towards understanding} the trade{-}off between reliability and delays that might lead to packet drops due to the deadline-constrained bursty traffic with retransmissions, {a scenario of a wireless network utilizing a slotted-ALOHA random-access channel} is investigated. The main focus is to reveal and investigate further the trade-off between the number of retransmissions and the packet deadline as a function of the arrival rate. Hence, we are able to determine numerically the optimal probability of transmissions and number of retransmissions, given the packet arrival rate and the packet deadline. The analysis of the system was done by means of discrete-time Markov chains. Two scenarios are studied{: \emph{i)}} the collision channel model {(in which a receiver can decode only when a single packet is transmitted)}, and {\emph{ii)}} the case {for which receivers have} multi-packet reception capabilit{ies}. {A p}erformance evaluation for a user with {different} transmit probability and number of retransmissions is conducted, {demonstrating} their impact on the average drop rate and throughput{, while at the same time} showing {that there exists a set of} values{,} under which improved performance can be acquired.
\end{abstract}

\IEEEpeerreviewmaketitle


%
%
\section{Introduction}

Future wireless communication networks are envisioned to play a major role towards enabling autonomous systems in the context of the Internet of Things (IoT), comprising connected vehicles, smart devices, or fully automated factories; see, for example, \cite{Kalle:2018, nkenyereye2021network, wu2021network, ssimbwa2022jcn}. The data traffic produced from these wireless devices, referred to as machine-to-machine (M2M) communication, will significantly differ from the wireless traffic served by currently deployed wireless networks. In greater detail, wireless devices might transmit packets consisting of few bytes of information, while being sporadically active. Moreover, a massive number of devices may demand ubiquitous connectivity, and transmitting packets with extremely stringent latency and reliability requirements, as it is the case of mission critical M2M applications, supporting real-time closed-loop control, one of the essential mechanisms enabling such emerging applications \cite{2019:ComMag, wu2020cm, choi2021jcn}.

The rapid blooming of applications requiring deadline-constrained packet transmissions and multimedia broadcasting over wireless communication networks stimulated research on deadline-constrained broadcasting, relying on scheduling \cite{Master2016,MaoKoksalTON2016, FountoulakisWiOpt2017, hou2020jcn} and random access \cite{BaeCommL2013, BaeCommL2015Jun, BaeCommL2015Oct,BaeCommL2017, choi2020jcn}. The work in \cite{BaeCommL2013} obtained the optimal access probability of secondary nodes in a cognitive radio network towards maximizing the successful delivery probability (SDP) under specific deadline constraints using simple slotted-ALOHA. The issue of improving the reliability for the deadline-constrained one-hop broadcasting, based on the slotted-ALOHA with retransmission was investigated in \cite{BaeCommL2015Jun}. More specifically, the SDP was derived, as well as the optimal access probability for SDP maximization. Regarding retransmissions, their optimal value under specific throughput requirements was determined. Queuing analysis of deadline-constrained broadcasting, but without retransmission was investigated in \cite{BaeCommL2015Oct}. By modeling the system as a discrete-time {\texttt{Geo/Geo/1}} queue with a specific delivery deadline, several performance metrics were investigated, including the loss probability, queue length distribution, mean waiting time, and SDP. Nevertheless, the analysis of deadline-constrained broadcasting with retransmission{s has} not {been} analyzed {yet}. Furthermore, the paper in \cite{BaeCommL2017} studied a slotted-ALOHA network consisting of nodes with energy harvesting capabilities. For this setup, the author proposed an approximate analytical model for deriving the timely-delivery ratio, since the interaction of energy and data queues poses significant difficulties in obtaining an exact analytical model. Finally, the author in \cite{choi2020jcn} adopted exploration for multi{-}channel ALOHA through preambles before transmitting data packets in machine-type communication (MTC), showing a maximum throughput improvement by a factor of $2 - e^{-1} \approx 1.632$. Also, a steady-state analysis with fast retrial was performed, highlighting that the delay outage probability is significantly reduced for a lightly loaded system.

The findings of this study can be useful in practical code domain-based MPR, such as code-division multiple access (CDMA) and sparse code division multiple access (SCMA) systems, supporting deadline-constrained applications. At the same time, this paper can serve as a building block towards investigating packet scheduling in random access cooperative networks with deadline constraints. 
Overall, the contributions of the paper are as follows.  
\begin{list4}
\item First, the successful transmission probability is obtained for the case where a new packet is generated after the successful transmission of the previous one has been completed or that packet has been dropped, either due to reaching the maximum number of allowed retransmissions or expiration. In this part of the analysis, no data buffering is considered.
\item The second part of the analysis investigates stochastic bursty traffic with buffer-aided devices by means of a discrete-time Markov chain (DTMC) for single- and multi-packet reception (MPR). Two cases are distinguished: 
\begin{list4}
\item[\emph{(i)}] first, the case where a packet keeps being retransmitted until expiration or successful transmission. This case resembles that of \cite{BaeCommL2015Oct}, however, a different DTMC is constructed providing directly, the SDP; 
\item[\emph{(ii)}] second, the case where the deadline value is larger than the number of allowed retransmissions. A similar construction of the DTMC is used and, as a result, the system performance is analyzed. 
\end{list4}
\item Simulation results are included, validating the theoretical findings and demonstrating the effect of {different} transmit probabilit{ies} and the number of allowed retransmissions on the drop rate and the average throughput. Finally, the positive effect of {MPR} in the network{, as expected,} is highlighted.
\end{list4}


The remainder of the paper is structured as follows. Section~\ref{sec:model} presents the system model adopted in this study and the necessary preliminaries. Then, Sections~\ref{sec:analysis} and \ref{sec:analysis2} present the theoretical framework and analysis for the SDP when a packet is generated after the previous one was either successfully transmitted or dropped. Furthermore, the impact of bursty traffic and buffering is also analyzed. In Section~\ref{sec:numerical}, the numerical and simulation results are presented, giving several insights about the performance of the considered scenario. Finally, in Section~\ref{sec:conclusions} we draw conclusions and discuss possible extensions and future research directions.

%
%
%
%
\section{System Model and Preliminaries} \label{sec:model}

{In this section, the system model and the necessary preliminaries for the development of our study are presented. For convenience, Table~\ref{notation} includes the notation used throughout the paper.}

\begin{table}[ht]{
\centering
\caption{Summary of notation}
\label{table:1}
\begin{tabular}[t]{  m{3.5em} | m{19em}  } 
  \hline
  \textbf{Symbol} &  \\
  \hline
  $N$ & No. of nodes in the network \\ 
  $q$ & Transmission probability \\ 
   $D$ & Packet deadline \\ 
  $n$ & No. of retransmissions \\ 
  $\gamma$ & Signal-to-Interference-and-Noise Ratio (SINR) threshold \\ 
  $P_{tx}$ & Transmit power \\ 
  $P_{rx}$ & Received power \\ 
  $h$ & Small-scale fading random value (RV) \\ 
  $s$ & Received power factor \\ 
  $\alpha$ & Path loss exponent \\ 
  $r_i$ & Distance between node $i$ and the receiver \\ 
  $v$ & Rayleigh fading RV parameter \\ 
  $\eta$ & Noise power at the receiver \\ 
  $q$ & Transmission probability \\ 
  $p$ & Success transmission probability \\ 
  $\mathcal{T}$ & Set of transmitting nodes \\ 
  $c$ & No. of transmitting nodes \\ 
  $\lambda$ & Average probability of the packet arrival process \\ 
  $\mu$ & Success transmission probability of a packet at the head of the queue \\ 
  $\nu$ & Ratio of $q$ over $\mu$ \\
  $b$ & No. of backlogged nodes \\ 
  $p_{i,c-1}$ & Success transmission probability for node $i$ when $c$ nodes simultaneously transmit\\ 
  $p_s(n,D)$ & Success transmission probability before packet expiration  or  before  reaching the allowed no. of retransmissions \\ 
  $S$ & Successful transmission event \\ 
  $X$ & First transmission attempt event \\ 
  $U$ & Unsuccessful transmission event \\ 
  $M$ & Transition probability matrix \\ 
  $\pi(s)$ & Steady-state of state $s$ belonging  in  the  set of  states $\mathcal{S}$ from which a  successful transmission can take place \\
  $\pi(f_D)$ & Steady-state of state $f_D$ belonging  in  the  set of  states $\mathcal{F_D}$ transmission is the last \\
$\mathcal{F}$ & The rest of the states from which the last unsuccessful transmission of a packet can take place  \\ 
  \hline
\end{tabular}\label{notation}}
\end{table}

\subsection{Network Model}
In this work, a network comprising $N$ nodes, being in transmission range and sharing the same wireless channel is considered. While we can consider the case in which each node $i$ in the network might have its intended destination, in this paper, without loss of generality, we concentrate on the case where all the nodes transmit towards a common destination; this also motivates our work on multi-packet reception. Random access of the wireless medium is assumed and thus, each node is transmitting with probability $q_i$ (for simplicity of exposition, the same value of $q_i$ is assumed for all nodes, \ie $q_i=q~\forall i$). The time is slotted and each packet transmission requires one slot. Also, instantaneous and error-free acknow\-ledge\-ments/negative-acknow\-ledge\-ments (ACK/NACK) are transmitted by the receiver over a separate narrow-band channel.

Regarding the packet deadlines, when a new packet $j$ is generated, it has a deadline $D_j$ (for simplicity, it is assumed that $D_j$ is the same for all packets, \ie $D_j=D~\forall j$) and immediately enters the queue (in the case of bursty traffic). If packet deadline expires before the packet reaches the destination, then the packet is dropped from the queue. Here, two cases are considered for the number of allowed retransmissions: 
\begin{itemize}
\item[(a)] the case where each packet can be retransmitted until its expiration (\ie retransmitted up to $D-1$ times), and 
\item[(b)] the case where the number of retransmissions is equal to $n$, where $1 \leq n<D-1$. 
\end{itemize}
Retransmissions are necessary when the packet does not reach the destination, either due to a collision or unsuccessful packet reception at the receiver, as a result of the wireless nature of the channel. As a convention, the transmission of a packet from the queue is performed at the start of the slot and packets enter the queue at the end of the slot.

\subsection{Physical Layer Model}

It is considered that a packet from node $i$ is successfully transmitted to the receiver, if and only if the signal-to-interference-and-noise ratio (SINR) exceeds a certain threshold $\gamma_i$, called the capture ratio, \ie $\text{SINR}_i\geq \gamma_i$. Let $P_{\text{tx},i}$  be the transmit power of node $i$, and $r_i$  be the distance between node $i$ and the receiver. The received power, when node $i$ transmits, is $P_{\text{rx},i}=h_i s_i$, where $h_i$ is a random variable (RV) representing small-scale fading and $s_i$ is the received power factor. Under Rayleigh fading, $h_i$ is exponentially distributed \cite{Tse:2005}. The received power factor $s_i$ is given by $s_i = P_{\text{tx},i}r_i ^ {-\alpha}$, where $\alpha$ is the path loss (PL) exponent. When only node $i$ transmits, the success transmission probability for node $i$ is given by
\begin{align}
p_{i,0} = \exp \left(- \frac{\gamma_{i} \eta}{v_i s_i} \right),
\label{eq: probone} 
\end{align}
where $v_i$ denotes the Rayleigh fading RV parameter (\ie $h_i\sim \mathrm{Rayleigh}(v_i)$), and $\eta$ is the noise power at the receiver.

Let the set of transmitting nodes at any time slot be denoted by $\mathcal{T}$. When $c$ nodes transmit simultaneously ($c\coloneqq |\mathcal{T}|$), the successful transmission probability for node $i$ is given by \cite[Theorem~1]{nguyen2008optimization}
\begin{align}
\label{eq:succprobMPR}
p_{i,c-1}=\frac{\exp\left(-\frac{\gamma_{i}\eta}{v_i s_i}\right)}{\prod_{k\in \mathcal{T}\backslash \left\{i\right\}}{\left(1+\gamma_{i}\frac{v_k s_k}{v_i s_i}\right)}}.
\end{align}

{Here, in contrast to a basic slotted-ALOHA system exhibiting collisions, we assume that multiple simultaneous transmissions can be successfully performed.}

\subsection{Preliminary analysis}

The packet arrival process at a node $i$ is assumed to follow the Bernoulli distribution with an average probability $\lambda$. Let $\mu$ denote the probability that a packet at the head of the queue of node $i$ will be successfully transmitted  to the destination at a given time slot. Then, for the case when $b-1$ other nodes are backlogged (where $b\leq N$), and only node $i$ transmits, $\mu$ is given by $ \mu_1$, where $\mu_1$ is 
\begin{align}
\mu_1=p_{i,0}q (1-q)^{b-1}.
\end{align}
When $b-1$ other nodes are backlogged, and apart from node $i$, $c-1$ specific other nodes transmit ($2\leq c\leq b$), then $\mu$ is given by $\mu_c^{1}$, where $\mu_c^1$ is 
\begin{align}
\mu_{c}^{1}=p_{i,c-1}q^{c}(1-q)^{b-c}.
\end{align}
If we consider the case of any $c-1$ nodes out of the $b-1$ other nodes to transmit, then $\mu$ is given by $\mu_c$, where $\mu_c$ is 
\begin{align}
\mu_{c}={b-1 \choose c-1}p_{i,c-1}q^{c}(1-q)^{b-c}.
\end{align}
Combining all the cases together,  the probability $\mu$ that a packet at the head of the queue of node $i$ will be successfully transmitted to the destination in a given time slot is
\begin{align}
\mu\coloneqq \sum_{k=1}^{b}\mu_{k} = \sum_{k=1}^{b} {b-1 \choose k-1} p_{i,k-1}q^{k}(1-q)^{b-k},
\end{align}
where $k$ denotes the number of nodes transmitting simultaneously, including node $i$.

\section{Theoretical Analysis}
{This section provides the theoretical analysis for the SDP and the effect of bursty traffic and buffering.}

\subsection{SDP}\label{sec:analysis}

In this part, the SDP analysis of a packet being at the head of the queue is given. Here, a new packet is generated after the previous one has been successfully delivered or dropped. Note that a packet might be dropped either due to expiration or exceeding the number of allowed retransmissions $n$. Two cases are considered: 1) when $n=0$ {(i.e., one transmission is allowed)} and 2) when $0<n \leq D-1$ {(the number of total transmissions are more than one but less than or equal to the deadline of the packet)}.

When the traffic is characterized by deadlines, an important metric is the SDP, given that the packet is at the head of the queue, $p_s(n,D)$, which is the probability that a packet will be successfully delivered to the destination prior to its expiration or surpassing the {total} number of {allowed} transmission attempts{, $m\equiv n+1$} .

\subsubsection{$n=0$}
First, the case {for which no} retransmissions {are allowed} is considered{, i.e.,} when the allowed number of transmission attempts{,} $m${,} is $m=1$. Then, $p_s({n},D) {= p_s(0,D)}$ {and} becomes
\begin{align}\label{eq:ps1D}
p_s({0},D)&= \mu  + (1-{q})\mu+\ldots+(1-{q})^{D-1}\mu \nonumber \\
&=\mu\sum_{k=1}^{D}(1-{q})^{k-1}=\frac{\mu[ 1- (1-q)^D]}{q}.
\end{align}
Let $\nu\triangleq \mu/q$, then~\eqref{eq:ps1D} becomes
\begin{align}\label{eq:ps1Dnew}
p_s({0},D) 
=\nu[ 1- (1-q)^D].
\end{align}

\subsubsection{$0<n \leq D-1$}
Let $S$ denote the event that a packet will be successfully transmitted within the desired deadline, and let $X$ denote the event of the first transmission attempt. Then, $p_s({n},D)${, $1\leq n \leq D$,} is given by \cite{BaeCommL2015Jun}
\begin{align*}
p_s({n},D) = \sum_{k=1}^{D}\mathbb{P}(S,{n}|X=k)\mathbb{P}(X=k),
\end{align*}
{where $\mathbb{P}(\cdot)$ denotes the probability of an event and $\mathbb{P}(\cdot|\cdot)$ denotes the conditional probability of an event. 
Hence}, the value of $p_s(m,D)$ can be computed {iteratively} by: 
\begin{align*} 
p_s({n},D) =& q\left[ \nu + \left( 1- \nu \right)p_s({n}-1,D-1)\right] \nonumber \\
& + (1-q)q \left[\nu + \left( 1- \nu\right)p_s({n}-1,D-2)  \right] \nonumber  \\
& + (1-q)^2q \left[ \nu + \left( 1- \nu \right)p_s({n}-1,D-3)  \right] \nonumber  \\
& + \ldots + (1-q)^{D-1}q \left[ \nu + \left( 1- \nu \right)p_s({n}-1,0) \right] \nonumber   \\
=& \sum_{k=1}^D (1-q)^{k-1}q \left[ \nu + \left( 1- \nu \right)p_s({n}-1,D-k)  \right].
\end{align*}

It has not been possible to derive a closed-form expression for $p_s({n},D)$; However, its value can be computed recursively.

%
%
%
%
\subsection{The effect of bursty traffic and buffering}\label{sec:analysis2}

In the previous section, we studied the case that a packet is directly generated once the previous one has been successfully transmitted or dropped. In this section, we consider the impact of bursty traffic and data buffering on the system performance. In greater detail, in a time slot, a packet will be generated with a given deadline, according to a probability $\lambda$ and will enter the queue. Our analysis focuses a single node having bursty traffic, while the rest $N-1$ nodes are considered to always be backlogged, \ie at least one packet always resides in their queues. It should be noted that this is the worst-case scenario, as the number of transmitting and consequently, the number of collisions in a time slot are overestimated.
A DTMC model of the system with bursty traffic is used, while the arriving packets are stored in a queue. Based on the number of allowed retransmissions, we consider two cases: 1) each packet can be retransmitted as many times as required for as long as it remains in the queue (\ie $n=D-1$ number of retransmissions); 2) a limited number of retransmissions (\ie $n<D-1$) is allowed for each packet.

\subsubsection{Number of retransmissions $n = D-1$}
\begin{figure}[t]
\centering
\includegraphics[width=\columnwidth]{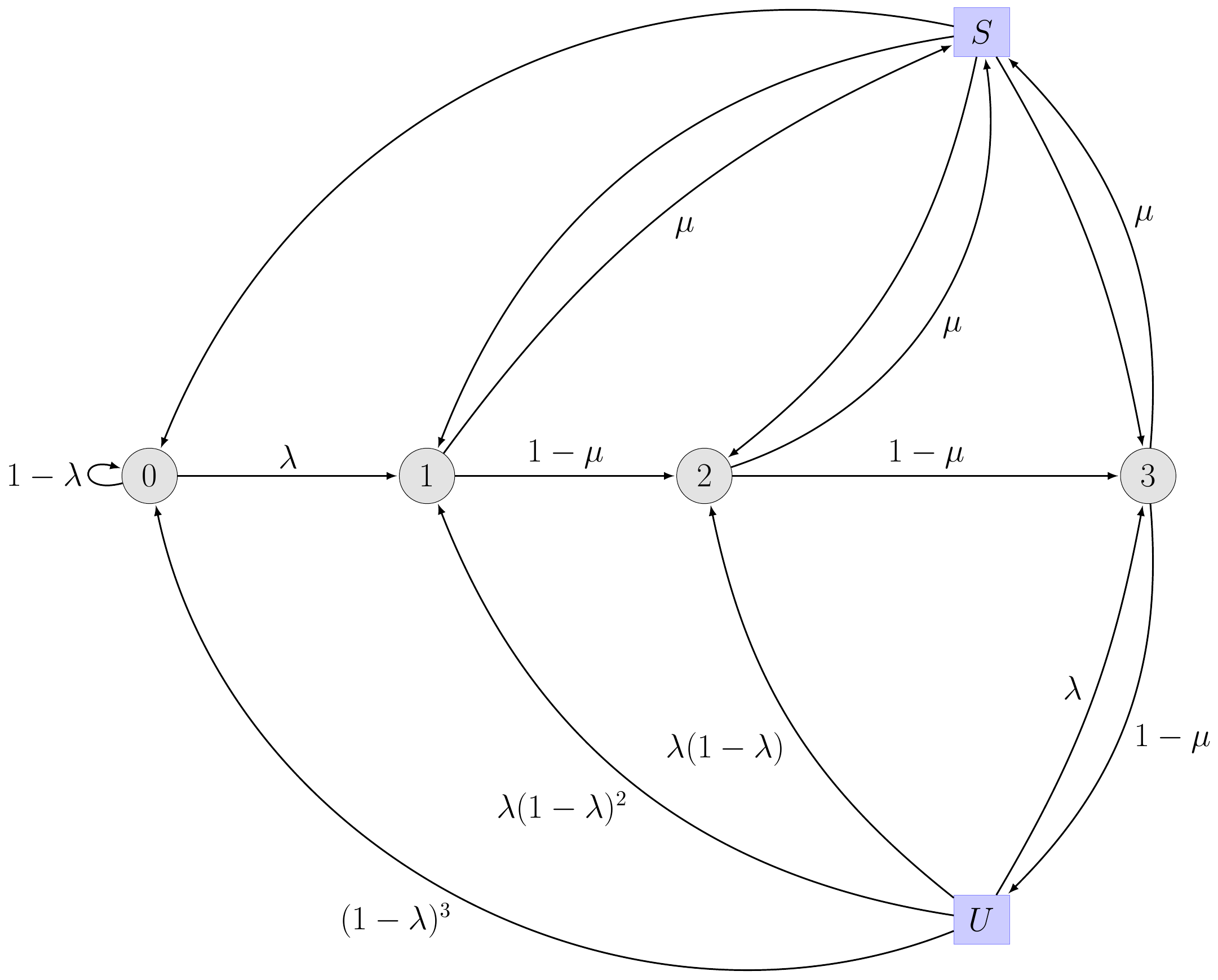}
\caption{DTMC when the number of allowed retransmissions is {$D = 3$, $n=2$}.}
\label{fig:MC1}
\end{figure}

Here, the number of possible retransmissions is equal to the deadline.
For a given deadline $D$, the DTMC has $D+1$ states, where the states from $0$ to $D$ model the passed time. The DTMC that he have constructed includes two additional \emph{virtual} states (\ie states $S$ and $U$) capturing the events of successful and unsuccessful delivery, respectively. State $S$ is included to capture the successful packet transmission before expiration or dropping. {The values of the transition probabilities from state $S$ to any other state depend on the state prior to $S$.} Then, state $U$, models the event where a packet either expires or it is dropped. An illustrative example of the DTMC for a deadline {$D=3$} is shown in {Fig.~}\ref{fig:MC1}.

The transition probability matrix $M$ for a deadline of {$D=3$}, which is column stochastic by construction, is given by:
{
\begin{align*}
M=\begin{bmatrix}
1- \lambda 	& \mu(1- \lambda) 				& \mu(1- \lambda)^2  		& \mu(1- \lambda)^3	\\
\lambda 		& \mu\lambda 					& \mu\lambda (1- \lambda) 	& \mu\lambda (1- \lambda)^2	 \\
0 			& 1-\mu 						& \mu\lambda 				& \mu\lambda (1- \lambda)	 \\
0 			& 0 							& 1-\mu 					& \mu\lambda	
\end{bmatrix} .
\end{align*}}

\subsubsection{Number of retransmissions $n < D-1$}

\begin{figure}[t]
\centering
\includegraphics[width=\columnwidth]{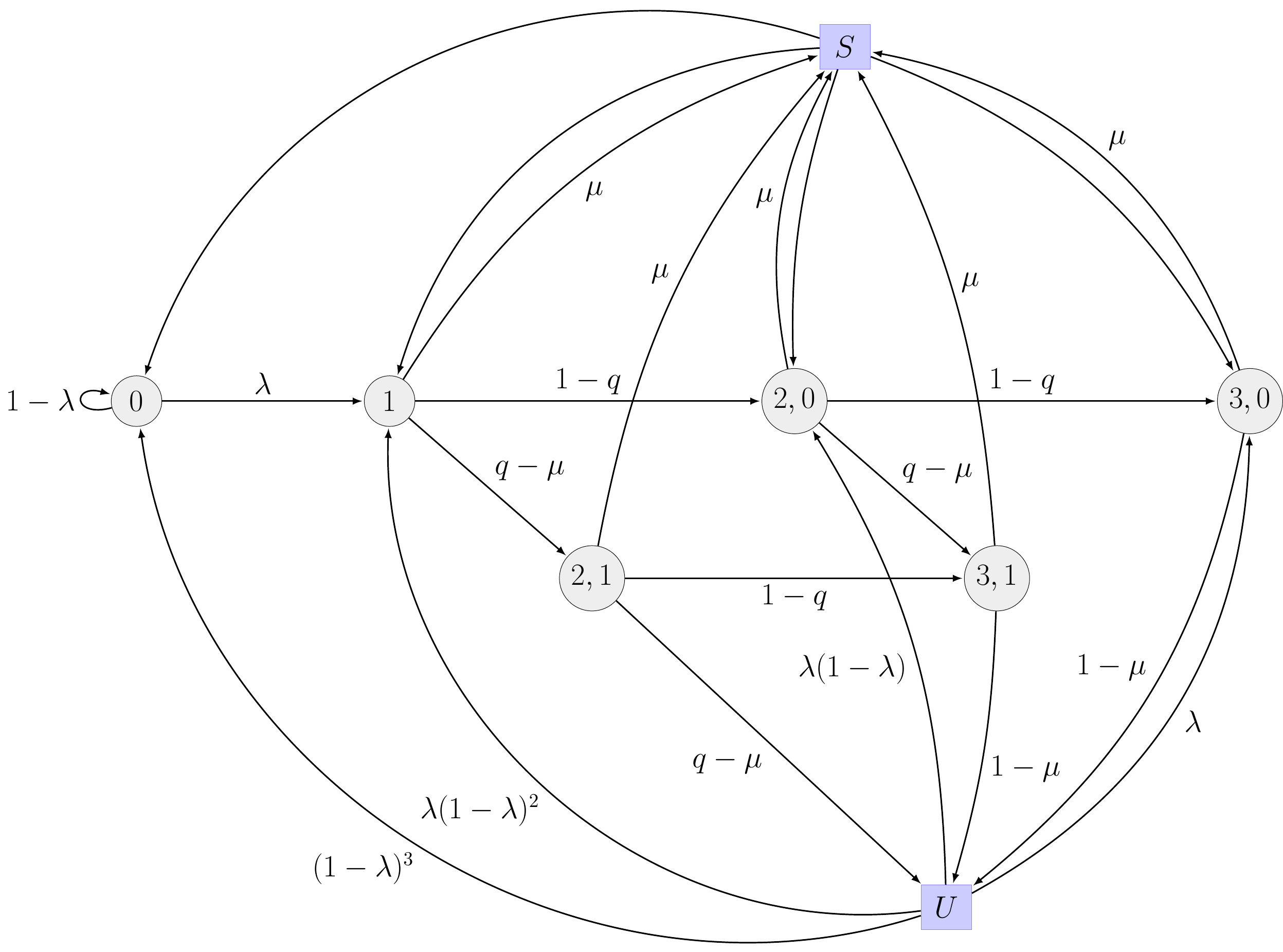}
\caption{DTMC for {$D=3$} when the number of allowed retransmissions is $n =1$.}
\label{fig:MC2}
\end{figure}

In this case, a packet drop might happen for two reasons: either the number of retransmissions has reached its maximum allowed value or the packet has expired. The main motive behind limiting the number of retransmissions is to reduce the number of packets residing at the queue and waiting to be transmitted, thus reducing the amount of packets that will expire, and increasing the system throughput.
Given a deadline $D$, the DTMC has {$D(n+1)+1 -\frac{n(n+1)}{2}$} states {including, similar to the previous case, two additional \emph{virtual} states, facilitating the visualization of the successful transmissions}. For the case which one retransmission is allowed ($n=1$), for example, the DTMC has $2D$ states (for a deadline of {$3$}, as illustrated in {Fig.~}\ref{fig:MC2}, {$6$} states exist).

\begin{remark}
Constructing these DTMCs that include these virtual states offers a framework for investigating the behavior of such advanced wireless communication systems where packets may have a limited number of allowed retransmissions.
\end{remark}

The throughput, $T$, of the system can be obtained by considering the probability of being at the virtual state $S$. This can be found by identifying the states from which a successful transmission can occur. In greater detail, the throughput is given by 
\begin{align}
T=\sum_{s\in \mathcal{S}} \pi (s) \mu,
\end{align}
where  $\pi (s)$ denotes the steady-state of state $s$ in the DTMC, $s$ being a state of the DTMC belonging in the set of states $\mathcal{S}$ capturing the event of a successful transmission. In the case of the DTMC in {Fig.~}\ref{fig:MC2}, set $\mathcal{S}$ consists of the states $\{1, (2,0), (2,1), (3,0), (3,1)\}$.  

The drop rate (or drop probability), $DR$, is correspondingly derived by considering the states from which a packet might be dropped, due to violating its deadline or absence of retransmissions. More specifically, the drop rate is given by
\begin{align}
DR = \sum_{f_D \in \mathcal{F}_D} \pi (f_D) (1-\mu)+\sum_{f\in \mathcal{F}}\pi(f) (q-\mu),
\end{align}
where $\pi(f_D)$ is the steady-state of the states $f_D$ in the DTMC, belonging is the set of states $\mathcal{F}_D$, whose transmission is the last. In the case of the DTMC in {Fig.~}\ref{fig:MC2}, $\mathcal{F}_D$ consists of the states $\{(3,0), (3,1)\}$. The remaining states from which the last unsuccessful transmission of a packet can take place belong to the set of states $\mathcal{F}$. In the case of the DTMC in {Fig.~}\ref{fig:MC2}, $\mathcal{F}$ consists of the state $\{ (2,1) \}$.

\begin{remark}
In order to find the optimal transmission probability $q$ with respect to (wrt) some metric (e.g., minimize $DR$ or maximize $T$), an optimization problem can be formulated. For example, for maximizing throughput $T$, we cast the following optimization problem:
\begin{align}
\max_{q}\sum_{s\in \mathcal{S}} \pi (s) \mu.
\end{align}
Note that both $\mu$ and the steady-state distribution $\pi$ are functions of $q$, but since there is no known analytical expression of $\pi$ wrt $q$ (yet), it is not possible to provide an analytical solution; hence, to solve this problem we resort to numerical approaches.
\end{remark}

\section{Simulation and Numerical Results}\label{sec:numerical}

This part presents simulation and numerical results{, using MATLAB\textsuperscript{\textregistered}} for a topology comprising buffer-aided nodes with buffer size\footnote{Our analysis does not consider the impact of varying the buffer size, as we assume that the buffer size is at least as big as the packet deadline value and in case the buffer size exceeds this value, the packets at the back of the queue will be never transmitted.} $L = 3$ and varying packet deadline values $D$. 
More specifically, the performance of a non-backlogged user is evaluated, in terms of drop rate and average throughput, measured in bps/Hz, for {10$^5$ time-slots,} various transmit probability $q$ and number of retransmissions $n$ values. It is noted that various cases are examined regarding the number of users in the network, \ie $N = 2, 3, 4, 5$, where apart from the non-backlogged user, the other users are assumed to be backlogged. 
Also, for the non-backlogged user, the packet arrival probability is $\lambda = 0.5${, unless otherwise stated} and the probability of a successful transmission is $p_{i,0} = 0.75$ for {the scenario with the collision channel model, while when MPR is allowed} $p_{i,1} = 0.375$, $p_{i,2} = 0.1875$, $p_{i,3} = 0.09375$ and $p_{i,4} = 0.046875$. {Taking into consideration eq.~(\ref{eq: probone}) and eq.~(\ref{eq:succprobMPR}), details of the simulation parameters for the considered network are given in Table~\ref{parameters}}.

\begin{table}[ht]{
\centering
\caption{Network parameters}
\begin{center}
\begin{tabular}{  l | r  } 
  \hline
  \textbf{Parameter} & \textbf{Value} \\
  \hline
  SINR threshold $\gamma_i$ & 0 dB \\ 
  Noise power $\eta$ & -115.4 dBm \\ 
  Transmit power $P_{tx_i}$ & 0.01 mW \\ 
  Rayleigh RV parameter $v_i$ & 1 \\ 
  Transmitter - receiver distance $r_i$ & 100 m \\ 
  Transmitter - receiver PL exponent $\alpha_i$ & 4.5 \\ 
  \hline
\end{tabular}\label{parameters}
\end{center}}
\end{table}

\subsection{{Collision channel model}}
The first scenario focuses on a topology with $N=2$ users, allowing only single-user transmission. So, when both users aim to access the channel, a collision occurs.

\subsubsection{Number of retransmissions $n=D-1$}

\begin{figure}[t]
\centering
\includegraphics[width=\columnwidth]{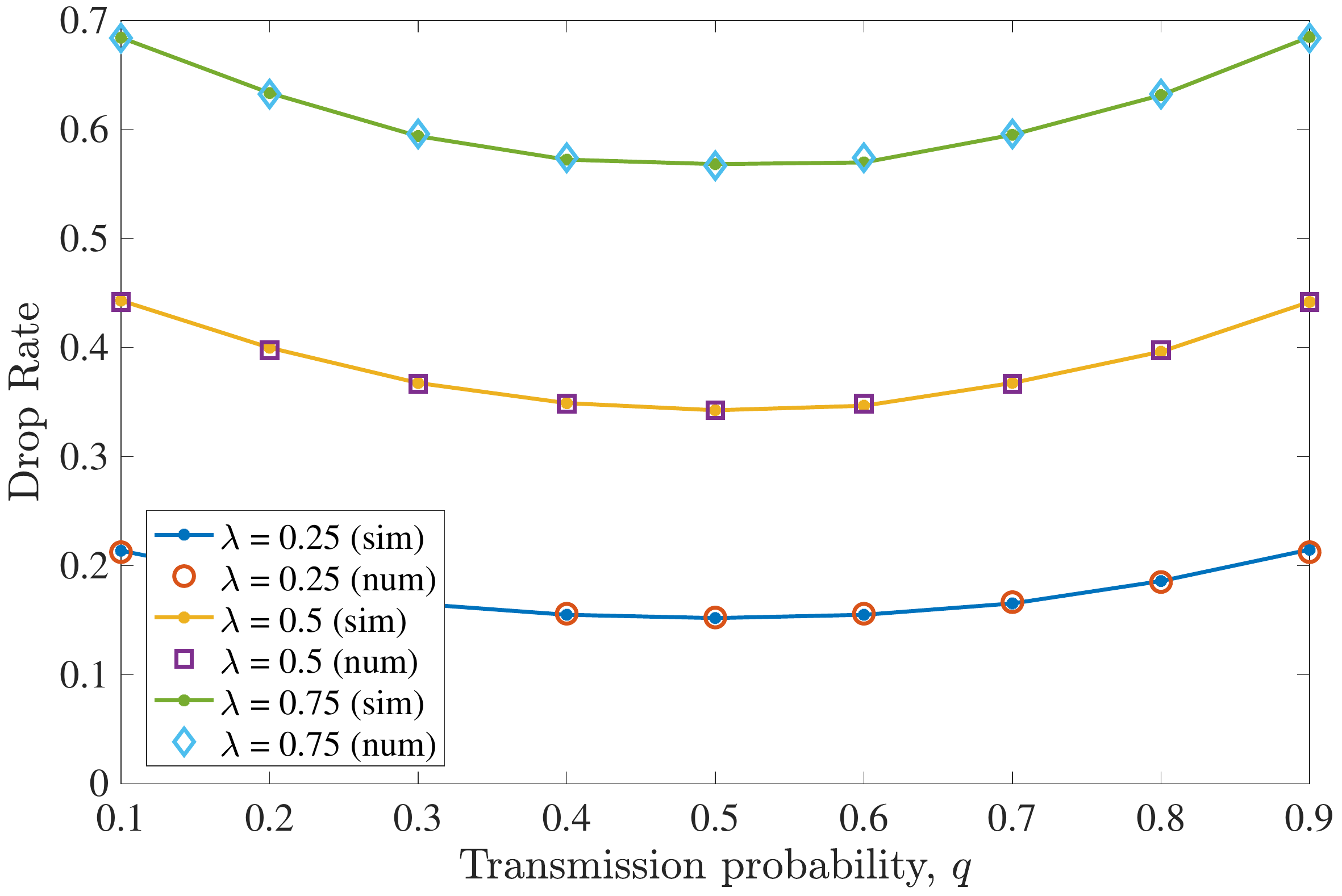}
\caption{Drop rate for various $q$ {and $\lambda$} values for the first DTMC for which $n=D-1=2$. A perfect match among the numerical and simulation results is observed. {Increasing $\lambda$ results in more packet drops.}}
\label{fig:drop1}
\end{figure}

The impact of varying $q$ for  $n = D-1= 2$ allowed retransmissions is considered in this scenario. {Fig.~}\ref{fig:drop1} shows the drop {rate} results for different values of transmit probability $q$, {equal} for the two users. It can be seen that the drop {rate} for the non-backlogged user is minimized for transmit probability value $q = 0.5$. For lower $q$ values, an increase in the drop {rate} is experienced, as the device does not access the frequently vacant wireless channel. {In these cases, packets reside for more time-slots in the queue, being consequently dropped due to expiration.} Moreover, for higher $q$ values, collisions may occur and the drop {rate} increases. Regarding the effect of $\lambda$, increasing the packet arrival probability results in increased drop rate.

\begin{figure}[t]
\centering
\includegraphics[width=\columnwidth]{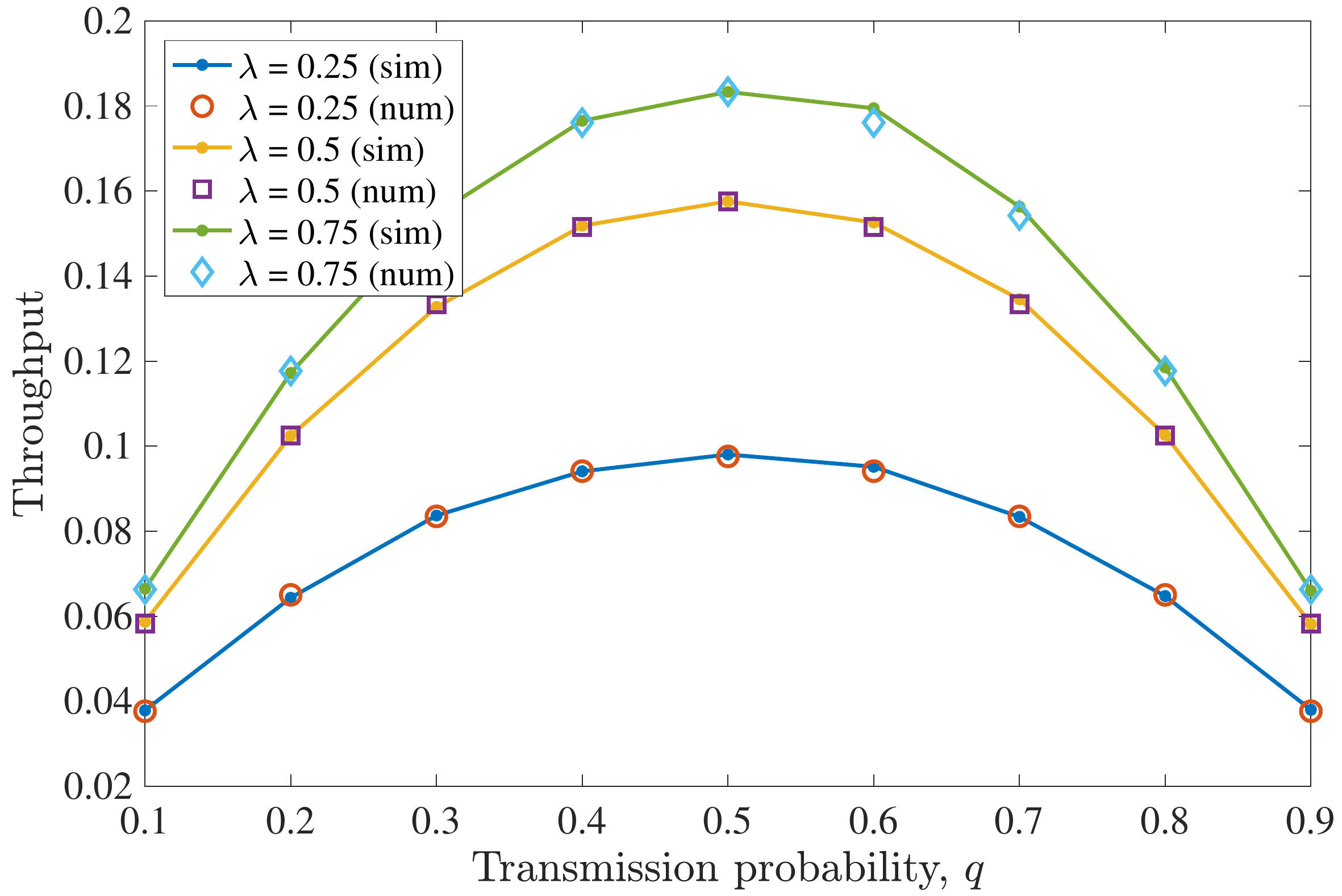}
\caption{Average throughput for various $q$ {and $\lambda$} values for the first DTMC for which $n=D-1=2$. Throughput is maximized when $q = 0.5$. {As $\lambda$ increases, throughput increases and a saturation is observed.}} 
\label{fig:thr1}
\end{figure}

Then, {Fig.~}\ref{fig:thr1} illustrates the average throughput performance for different $q$ values. Here, maximum throughput is obtained when $q = 0.5$. As it has been already observed for the drop {rate} results, {when this $q$ value is adopted, the non-backlogged user enjoys improved throughput, as more packets are successfully transmitted towards the destination. More specifically, a $q = 0.5$ optimizes the trade-off among channel access and collisions with the packets of the backlogged user. {Also, increasing $\lambda$ results in increased throughput for the non-backlogged user, while a saturation can be seen when varying $\lambda = 0.5$ to $\lambda = 0.75$.}}  

\subsubsection{Number of retransmissions $n<D-1$}

\begin{figure}[t]
\centering
\includegraphics[width=\columnwidth]{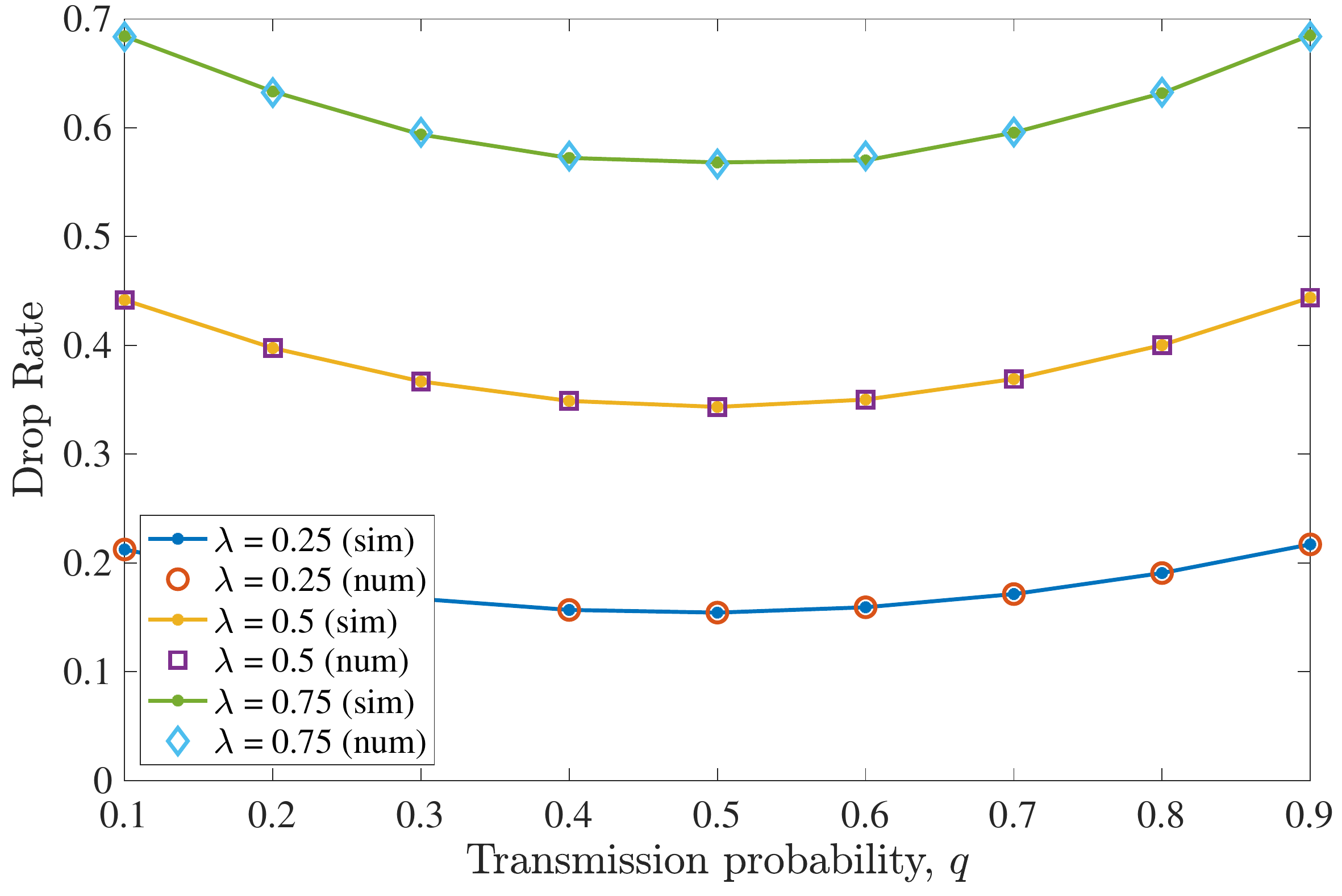}
\caption{Drop rate for various $q$ {and $\lambda$} values for the second DTMC for which $D=3$ and $n=1$. {Drop rate is slightly increased for  higher $q$, compared to the case of $ n = D-1$. {More packets are dropped when $\lambda$ increases.}}}
\label{fig:drop2}
\end{figure}

For this scenario, the maximum number of allowed retransmissions in the network is $n = 1$. {Fig.~}\ref{fig:drop2} shows the drop rate performance for different $q$ values. Again, a similar trend with the first scenario is observed. More specifically, for the non-backlogged user, the minimum drop rate is seen when $q = 0.5$. As for the effect of $n$ on the drop rate, a negligible increase is observed for larger $q$ values, since $n$ is equal to one and packets are dropped more often from the system, after a collision. {It must be noted that similar to the previous case, higher $\lambda$ values lead to more packets being dropped.}

\begin{figure}[t]
\centering
\includegraphics[width=\columnwidth]{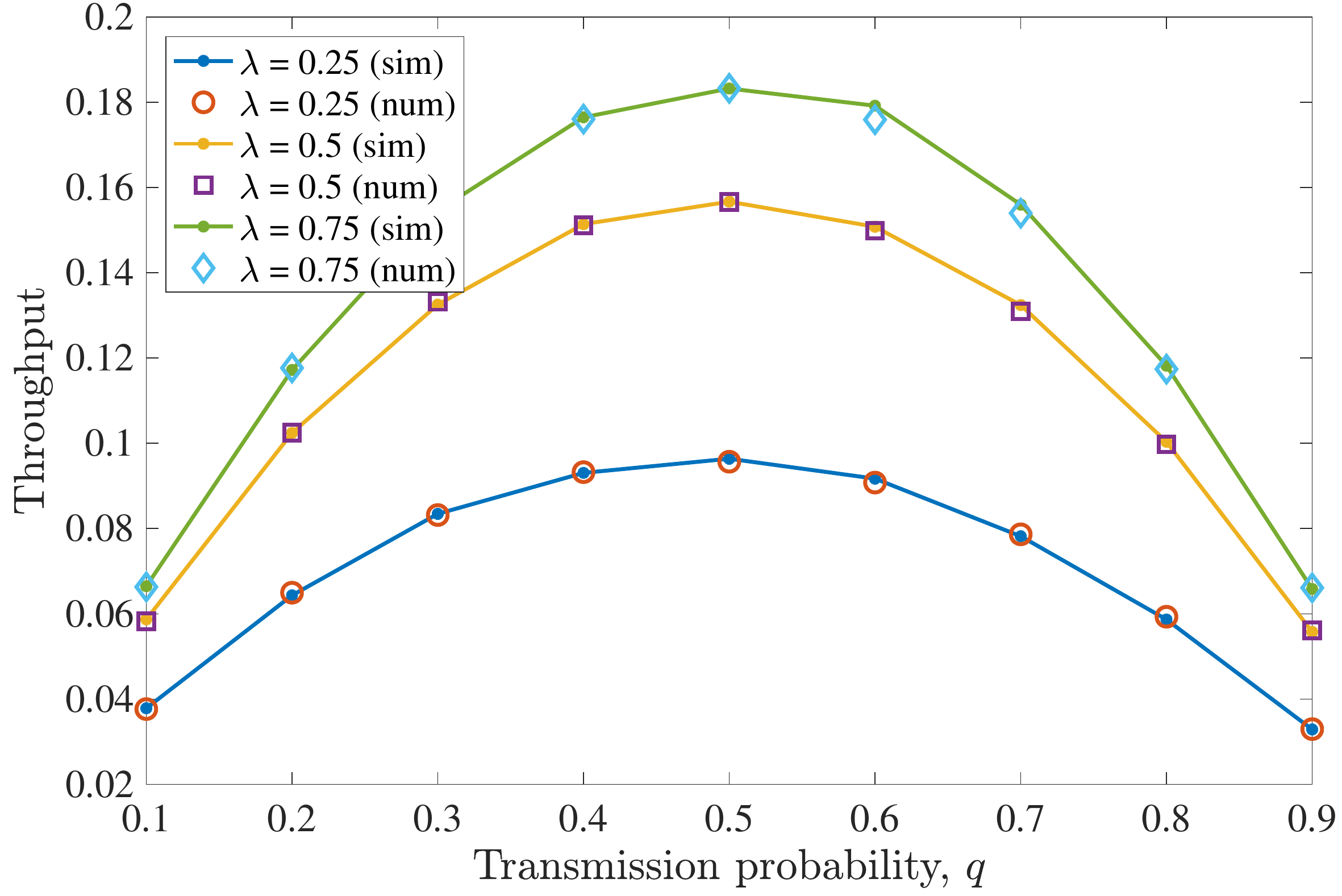}
\caption{Average throughput for various $q$ {and $\lambda$} values for the second DTMC for which $D=3$ and $n=1$. Throughput {slightly reduces}, compared to the case of $ n = D-1$ {and increases for higher $\lambda$ values}.}
\label{fig:thr2}
\end{figure}

Finally, {Fig.~}\ref{fig:thr2} depicts average throughput results for varying $q$ values. The throughput performance closely follows that of the first scenario. So, for $q = 0.5$ maximum throughput is acquired for the non-backlogged user, while for higher $q$ values, the average throughput slightly degrades. In conclusion, when both cases of $n$ are considered, it can be observed that the performance of the non-backlogged user does not change and the number of retransmissions marginally affect the drop rate and the average throughput given small deadline values and a low number of users. {Finally, throughput performance improves when higher $\lambda$ values characterize the non-backlogged user's traffic, resulting in saturation.}

\subsubsection{Comparisons}

\begin{figure}[t]
\centering
\includegraphics[width=\columnwidth]{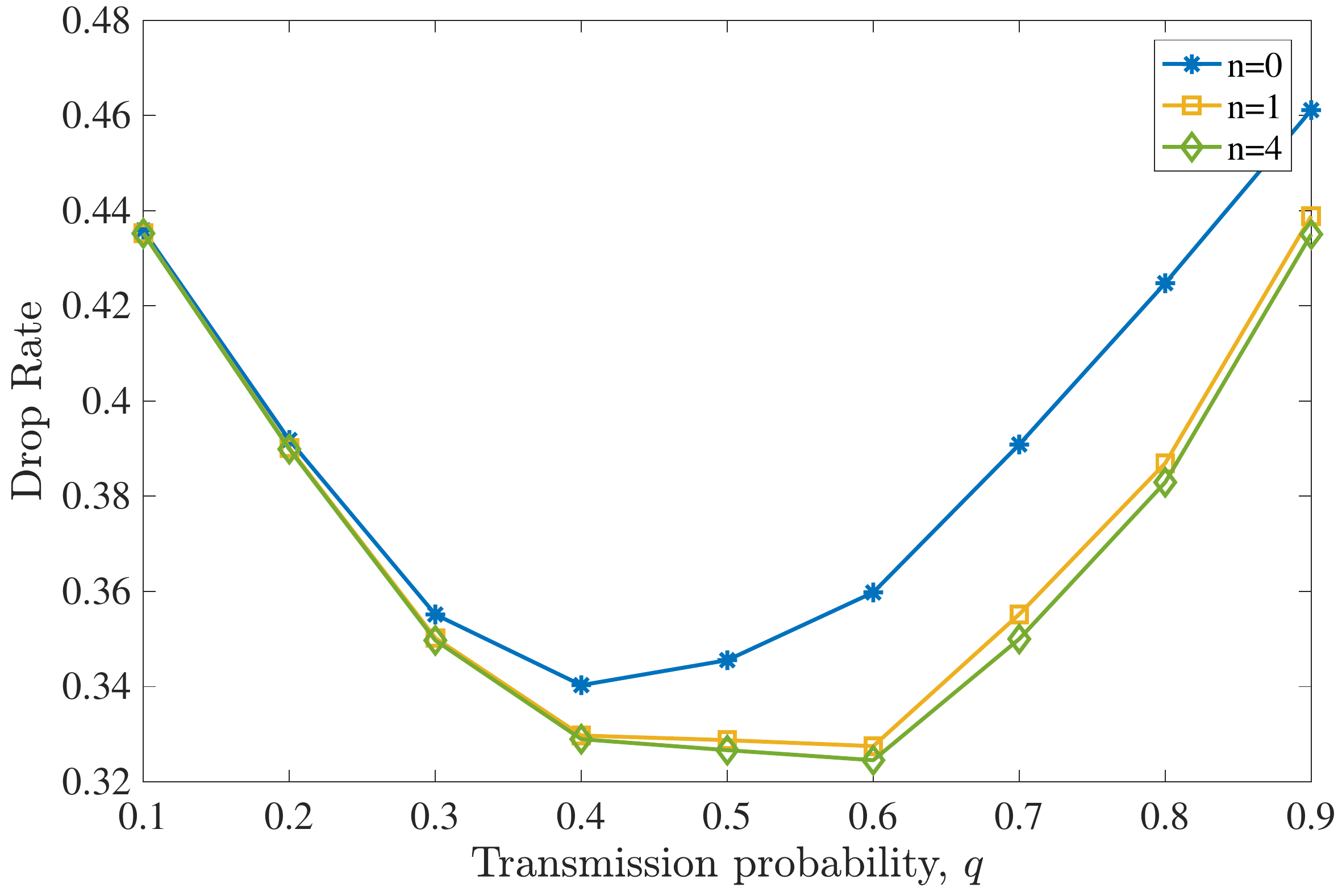}
\caption{Drop rate for various $q$ values for $D=5$ and $n=0, 1, 4$. {Drop rate performance improves for increasing $n$. Allowing $n = 4$ retransmissions offers a small drop rate reduction, compared to the case of $n = 1$.}}
\label{fig:drop2diff}
\end{figure}

\begin{figure}[t]
\centering
\includegraphics[width=\columnwidth]{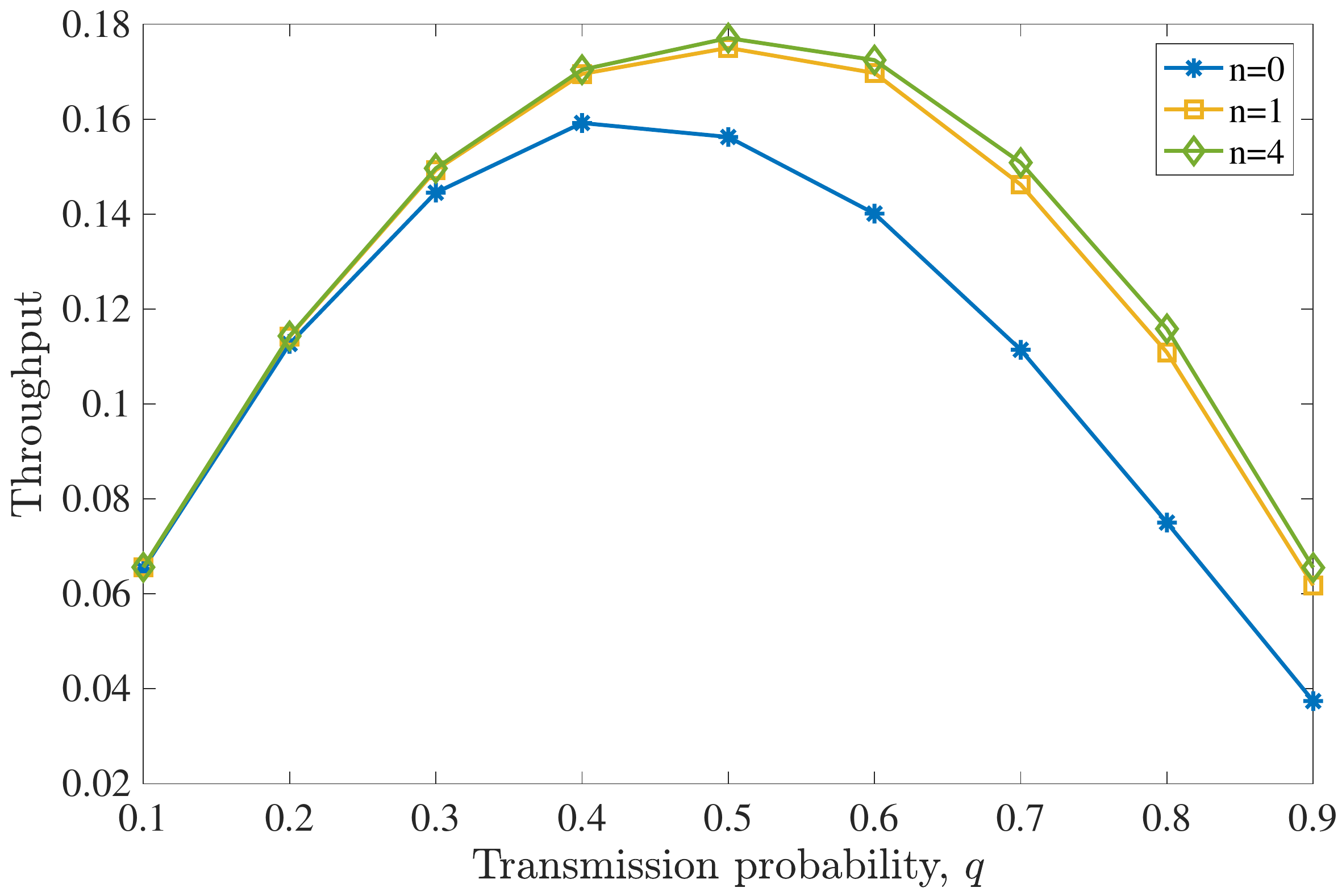}
\caption{Average throughput for various $q$ values for $D=5$ and $n=0, 1, 4$. {Throughput is enhanced for increasing $n$. However, only a slight improvement is observed when $n = 4$ is adopted over $n = 1$.}}
\label{fig:thr2diff}
\end{figure}

In this case, the system performance is compared, when all the packets are characterized by deadline values$D=5$, and additionally, \emph{(i)} no retransmissions are attempted($n=0$), \emph{(ii)} some retransmissions are attempted (in this case $n=1$), and \emph{(iii)} as many retransmissions as needed are allowed ($n=D-1=4$). We see in both {Fig.~}\ref{fig:drop2diff} and {Fig.~}\ref{fig:thr2diff} that the performance is improved, as more retransmissions attempts are allowed. This is expected, as in this case, there are no packets with the same or lower deadlines residing further back in the queue, due to the fact that the inflow of packet is low and equal deadline is assumed to for all the packets.

\subsection{{Multi-packet reception}}
The second scenario investigates the impact of allowing MPR on the drop rate and throughput performance of the network. Here, the number of users $N$ varies, \ie $N = 2$ or $N = 3$ and $D=3$.

\subsubsection{Number of retransmissions $n=D-1$}
\begin{figure}[t]
\centering
\includegraphics[width=\columnwidth]{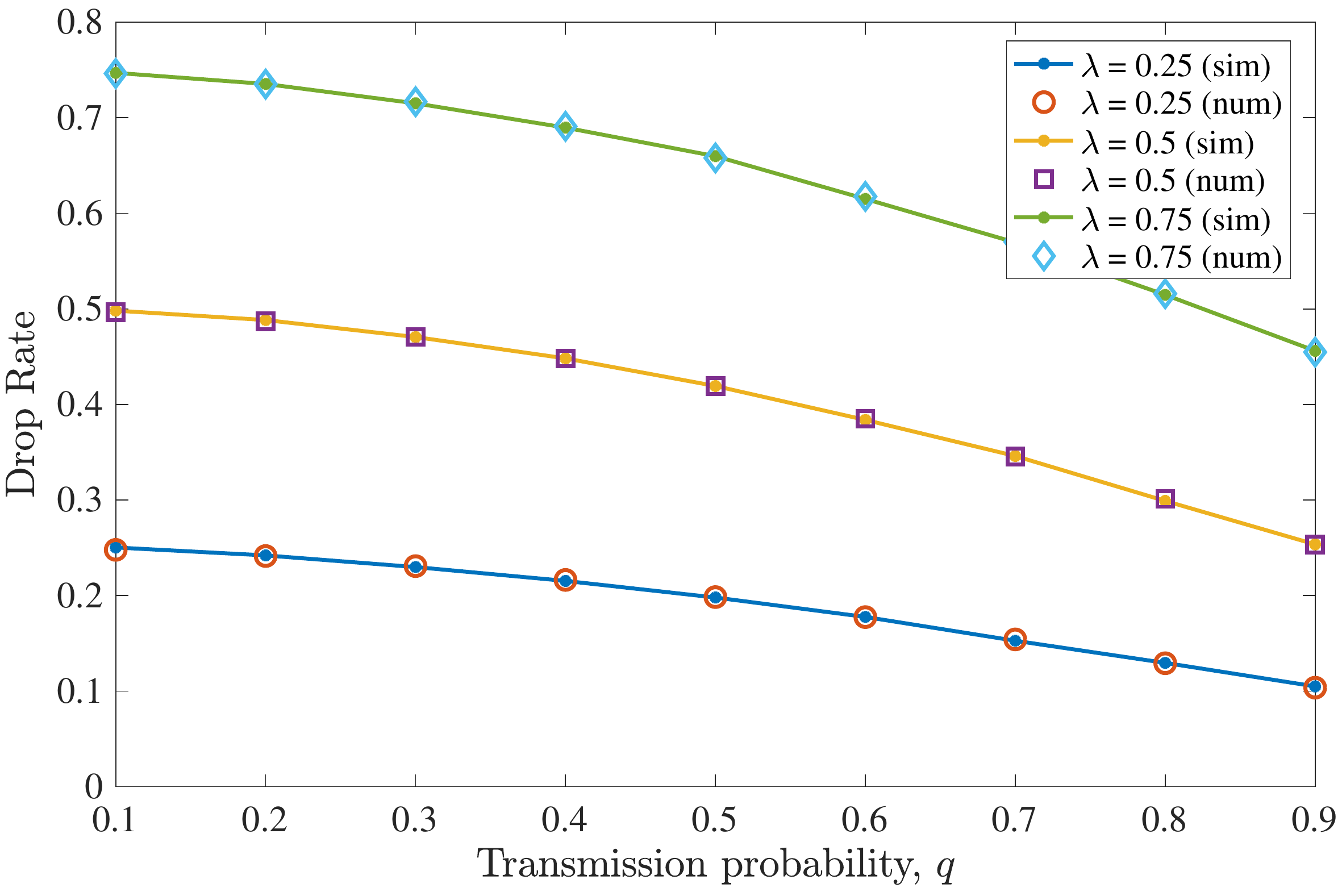}
\caption{Drop rate for various $q$ {and $\lambda$} values for the first DTMC for which $n=D-1=2$ and $N=c=2$. {It is observed that drop rate reduces as $q$ increases when MPR is allowed.} {Higher $\lambda$ values have a negative impact on the drop rate, independently of $q$.}}
\label{fig:drop1multi}
\end{figure}

{Fig.~}\ref{fig:drop1multi} depicts the drop rate performance when $n=2$ retransmissions are allowed in a network with $N=2$ users. As both users simultaneously transmit, significantly different system behavior is observed, compared to the previous scenario due to the MPR capability. In greater detail, considering the fixed value of the successful transmission probability $P$, it can be observed that the drop rate reduces as $q$ increases as MPR is allowed, \ie collisions do not occur in the network. {In addition, increasing $\lambda$ results in more instances of dropped packets regardless of $q$.}

\begin{figure}[t]
\centering
\includegraphics[width=\columnwidth]{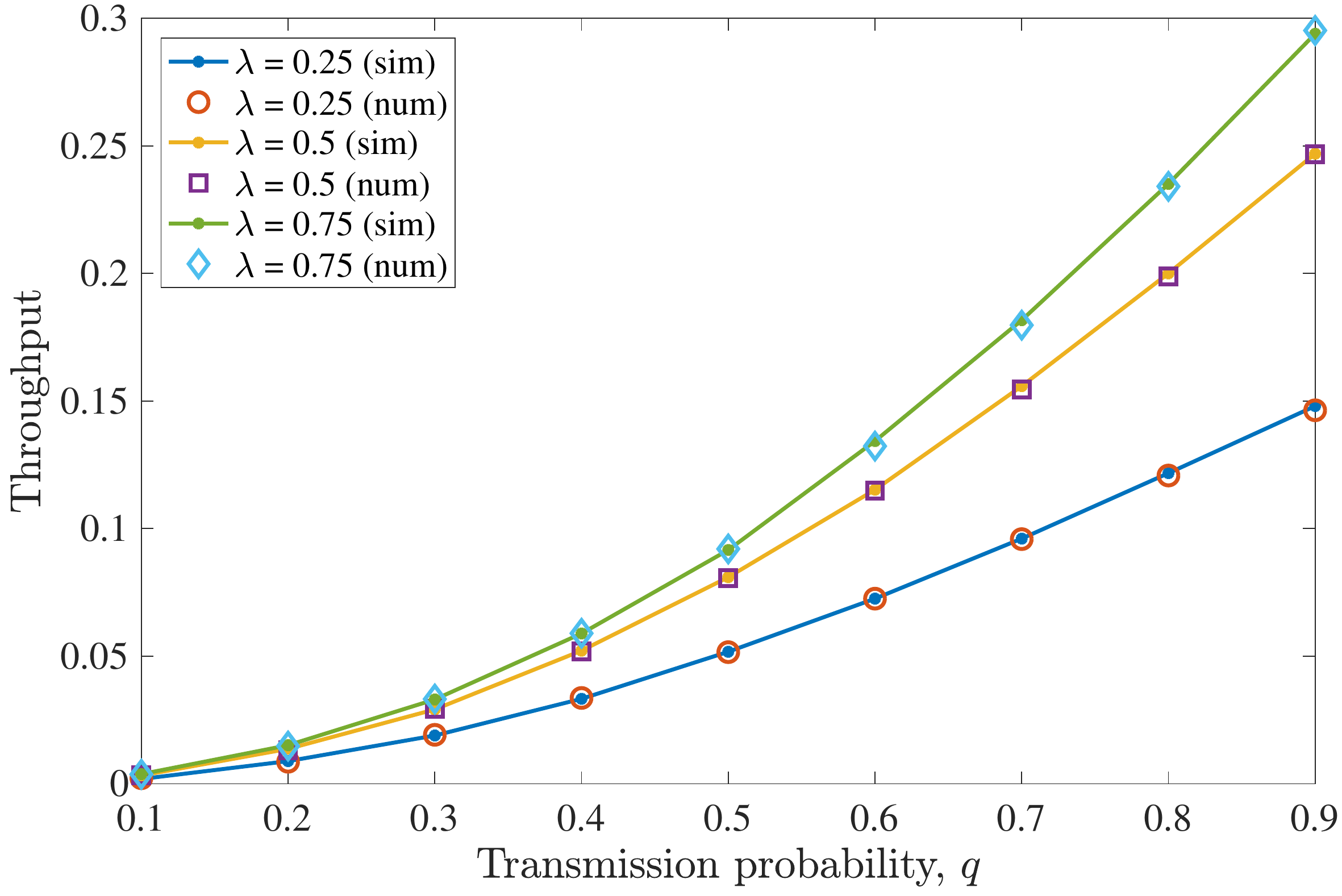}
\caption{Average throughput for various $q$ {and $\lambda$} values for the first DTMC for which $n=D-1=2$ and $N=c=2$. {As MPR is allowed, increasing $q$ has a beneficial impact on throughput.} {For increased $\lambda$ values throughput is significantly increased for higher $q$.}}
\label{fig:thr1multi}
\end{figure}

Next, {Fig.~}\ref{fig:thr1multi} shows the throughput performance of the network. As it was observed for the drop rate, the average throughput performances improves when the value of $q$ increases. Since MPR is allowed, allowing both users to access the medium more frequently has a beneficial effect on the number of transmitted packets and thus, the average throughput is increased. {In this case, a higher $\lambda$ corresponds to significantly higher throughput as $q$ increases, while for lower $q$ values throughput is slightly affected.}

\subsubsection{Number of retransmissions $n<D-1$}

\begin{figure}[t]
\centering
\includegraphics[width=\columnwidth]{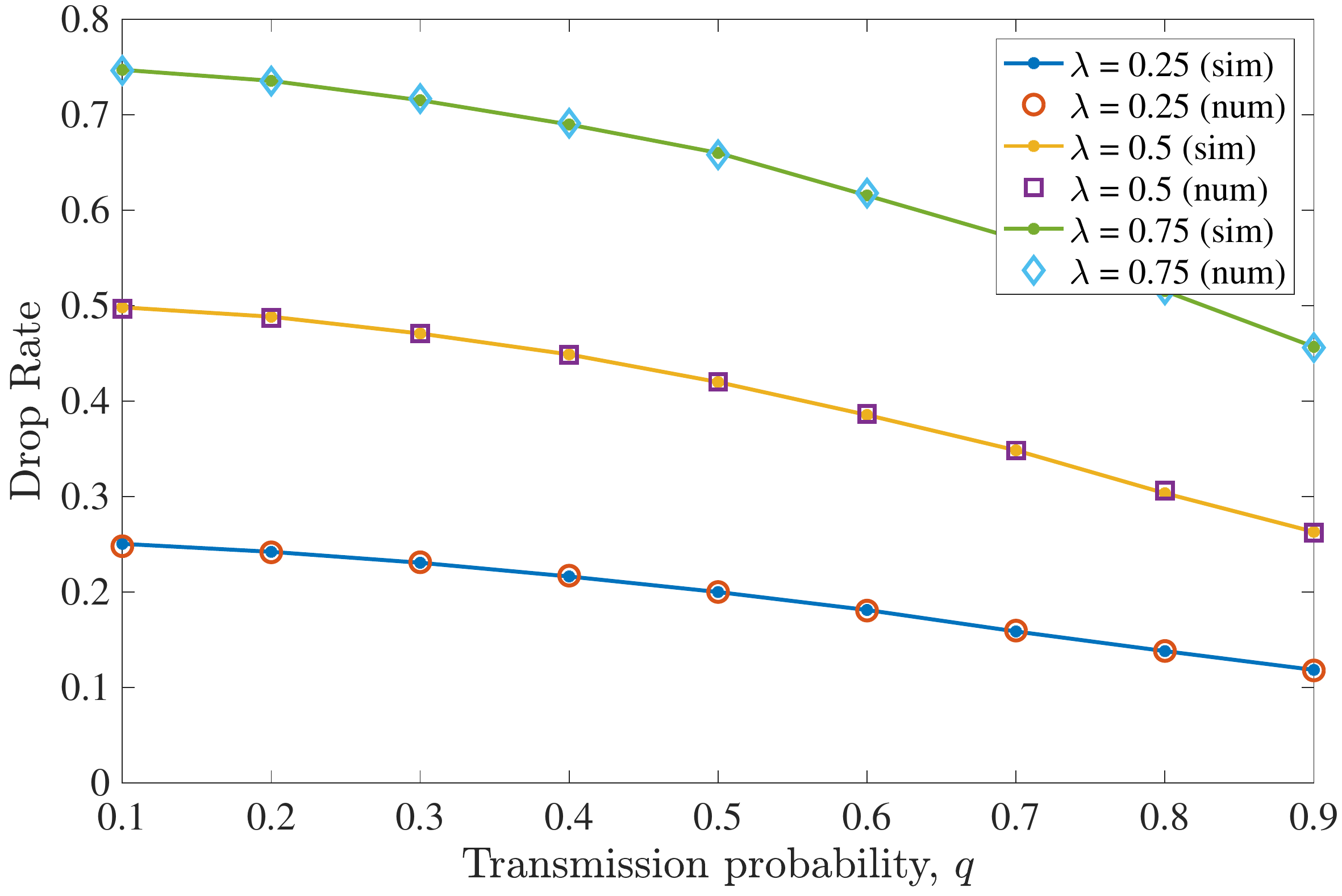}
\caption{Drop rate for various $q$ {and $\lambda$} values for the second DTMC for which $D=3$, $n=1$ and $N=c=2$. {Here, drop rate slightly increases compared to the case of $n = D-1$.} {A higher $\lambda$ degrades the drop rate performance for all $q$ values.}}
\label{fig:drop2multi}
\end{figure}

{Fig.~}\ref{fig:drop2multi} illustrates the drop rate performance for the same network when $n=1$ retransmission is possible. It can be seen that the reduction of allowed retransmissions results in almost negligible drop rate performance degradation for the non-backlogged user. Overall, drop rate reduces in an identical fashion as in the case of $n=2$ retransmissions, as $q$ increases. {On the contrary, when $\lambda$ increases, drop rate performance degrades.}

\begin{figure}[t]
\centering
\includegraphics[width=\columnwidth]{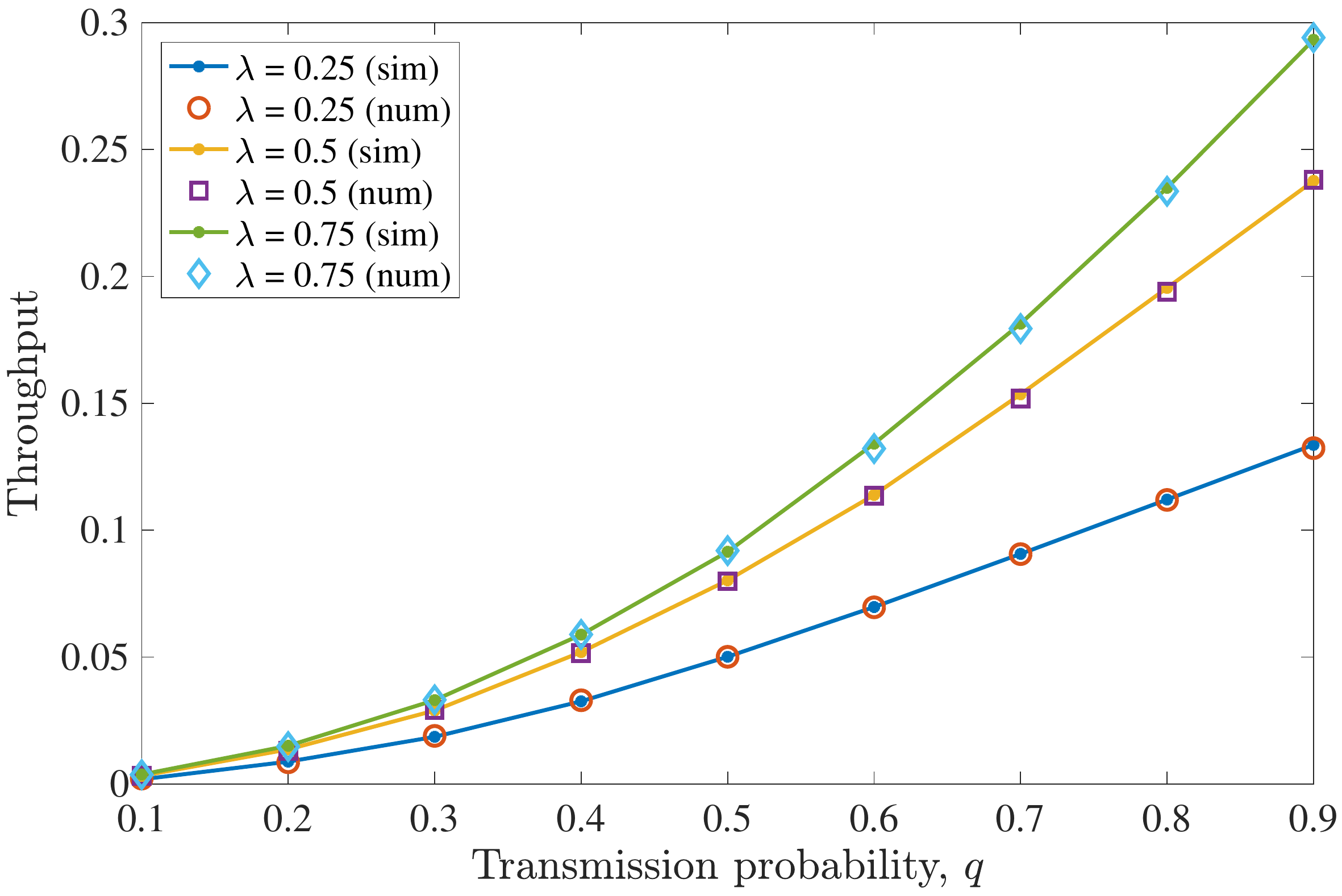}
\caption{Average throughput for various $q$ {and $\lambda$} values for the second DTMC for which $D=3$, $n=1$ and $N=c=2$. {The throughput performance is slightly worse compared to the case of $n=D-1$.} {As $\lambda$ increases, throughput performance improves for larger $q$ values.}}
\label{fig:thr2multi}
\end{figure}

Likewise, the average throughput performance is largely unaffected by reducing the value of $n$, as it can be seen in {Fig.~}\ref{fig:thr2multi}. More specifically, the average throughput sees a noteworthy increase at around 7$\%$ for $q = 0.9$, \ie when both users try to transmit with high probability. {In this comparison, when $\lambda = 0.25$, slightly improved throughput is observed compared to the corresponding case of $n = 2$ but for the other two $\lambda$ values throughput is almost negligibly reduced.}

\subsubsection{Comparisons}
Here, different cases with $N=2,3,4,5$ users are considered and the impact of varying $c$, \ie the number of simultaneously transmitting users on the drop rate and throughput performance is examined. 

{The drop rate results are depicted in {Fig.~}13. It can be seen that when $c=1$, the drop rate first reduces until $q=0.3$ for $N=2,3$ and until $q=0.2$ for $N=5$, since packet expiration due to inactivity is avoided. 
Then, as $c$ increases, worse drop rate performance is observed for all $N$ values. However, when all the users simultaneously transmit ($N=c$) and MPR is performed, the drop rate performance is enhanced, as $q$ increases, independently of the number of users in the network. Meanwhile, for this case, drop rate increases as $N$ increases due to the higher levels of interference in the network.}

\begin{figure}[t]
\centering
\includegraphics[width=\columnwidth]{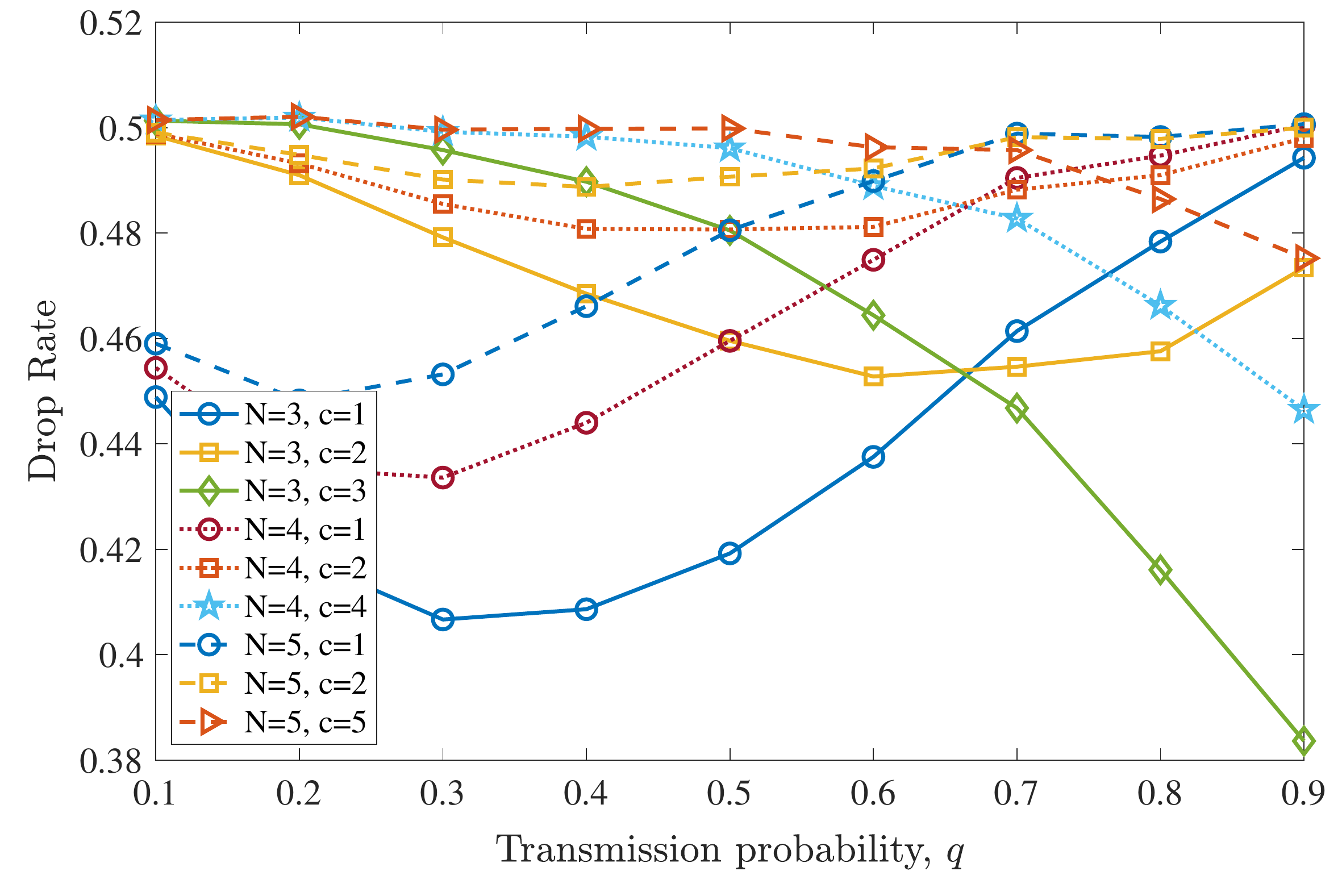}
\caption{{Drop rate for various $q$ values for the first DTMC for which $n=D-1=2$, $N=3, 4, 5$ and $c=1, \dots , N$. Significantly different performance is observed when $N>c$ compared to $N=c$ where drop rate reduces as $q$ increases.}}

\label{fig:drop2diffmulti}
\end{figure}

{Finally, {Fig.~}14 shows the average throughput results for the three different cases of $N$ and a varying number of concurrently transmitting users $c$. Again, results indicate that when MPR is allowed, throughput is improved when all the users are able to transmit and access the channel with higher probability. Moreover, increasing the number of users in the network $N$ negatively affects throughput, as interference arises. 
Also, for the case of $c=1$, a $q > 0.3$ for $N=3,4$ and a $q>0.2$ for $N=5$ leads to reduced throughput.}

It should be noted that even though, we investigate the network performance for a different number of users $N$, taking values {merely in the set} $N=\{2,\dots,5\}$, these users share a single wireless channel. However, in practical network deployments, as the available spectral resources increase, the number of coexisting users and machines on multiple channels will correspond to the massive connectivity cases of IoT and M2M communication scenarios.

\begin{figure}[t]
\centering
\includegraphics[width=\columnwidth]{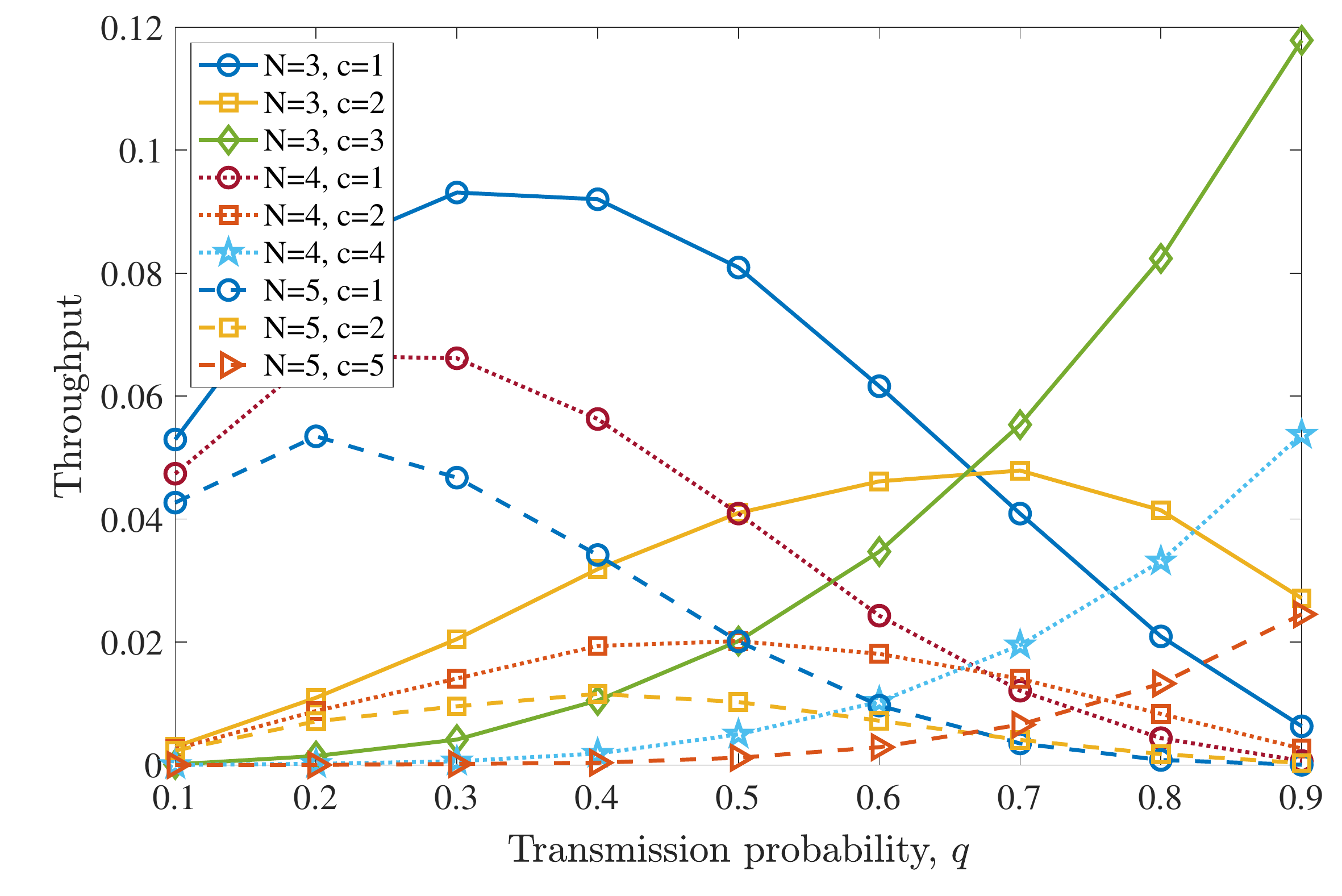}
\caption{{Average throughput for various $q$ values for the first DTMC for which $n=D-1=2$, $N=3, 4, 5$ and $c=1, \dots , N$. 
Throughput performance for $N>c$ is optimal for different $q$ values, while for $N=c$, an ever-increasing $q$ provides better throughput.}}

\label{fig:thr2diffmulti}
\end{figure}

\section{Summary and Future Directions}\label{sec:conclusions}

\subsection{Summary}

A network consisting of wireless devices transmitting deadline-constrained data using slotted-ALOHA random-access channels was considered. The performance of deadline-constrained transmission with retransmissions was studied, as the transmission of packets residing in queues might experience delays from retransmissions and they might get dropped before their turn comes for transmission. The goal of this study was to reveal and investigate the trade-off among the packet deadline and the number of allowed retransmissions, as a function of the packet arrival rate. 
Towards this end, the system was analyzed by using Markov chains, while numerical and simulation results highlighted the impact of different transmit probability values and the number of allowed retransmissions on the drop rate and the average throughput. Moreover, the effect of allowing multi-packet reception for different number of simultaneously transmitting users was shown.

\subsection{Future Directions}



There are several interesting research directions for extending this paper, in the context of MTC and 5G and beyond networks. Part of ongoing work includes the following directions: 
\begin{list4}
\item to investigate the performance of MPR using direct sequence CDMA (DS-CDMA) or SCMA for improving multiple access by allowing multiple simultaneous transmissions that would otherwise lead to a collision \cite{IoTJ_FD1}; 
\item to analyze the case of random access in cooperative networks with deadline constraints, departing from orthogonal schemes based on packet scheduling; see, e.g., \cite{IoTJ_FD2};
\item to find analytical methods to optimize the transmission probability $q$ for each scenario;
\item to take into consideration the freshness of the data \cite{IoTJ_FD3, fountoulakis2022jcn}; 
\item to consider power control mechanisms and study the interplay between improving the success transmission probability $p_{i,0}$ and number of transmissions \cite{Endrit:2020ICC}, and
\item to study the performance of random access in the context of grant-free non-orthogonal multiple access (NOMA) \cite{Abbas:tcom2018, choi2022wcm}.
\end{list4}

%
%
%
%

%
%
%
%

%
%
\bibliographystyle{IEEEtran}
\bibliography{biblio}

%
%
\end{document}